\documentclass[aps,prd,twocolumn,preprintnumbers,superscriptaddress,nobibnotes,floatfix,longbibliography,nofootinbib]{revtex4-1}

\pdfoutput=1
\usepackage{amsmath,amsfonts,amssymb,mathrsfs,graphicx,color,longtable}
\usepackage{hyperref}
\usepackage{graphicx}
\usepackage{amsfonts,amsmath,amssymb,bm}
\usepackage{array}
\usepackage{color}
\usepackage{enumitem}
\usepackage[explicit]{titlesec}
\usepackage[utf8]{inputenc}
\usepackage[normalem]{ulem}
\usepackage{xcolor}

\hypersetup{
     colorlinks   = true,
     citecolor    = red,
     urlcolor     = blue,
     linkcolor    = blue
}

\begin{document}
\title{New constraints on primordial features from the galaxy two-point correlation function}

\author{Mario Ballardini}\email{mario.ballardini@inaf.it}
\affiliation{Dipartimento di Fisica e Astronomia, Alma Mater Studiorum Universit\`a di Bologna, via Gobetti 93/2, I-40129 Bologna, Italy}
\affiliation{INAF/OAS Bologna, via Piero Gobetti 93/3, I-40129 Bologna, Italy}
\affiliation{INFN, Sezione di Bologna, viale C. Berti Pichat 6/2, I-40127 Bologna, Italy}
\affiliation{Department of Physics \& Astronomy, University of the Western Cape, Cape Town 7535, South Africa}

\author{Fabio Finelli}
\affiliation{INAF/OAS Bologna, via Piero Gobetti 93/3, I-40129 Bologna, Italy}
\affiliation{INFN, Sezione di Bologna, viale C. Berti Pichat 6/2, I-40127 Bologna, Italy}

\author{Federico Marulli}
\affiliation{Dipartimento di Fisica e Astronomia, Alma Mater Studiorum Universit\`a di Bologna, via Gobetti 93/2, I-40129 Bologna, Italy}
\affiliation{INAF/OAS Bologna, via Piero Gobetti 93/3, I-40129 Bologna, Italy}
\affiliation{INFN, Sezione di Bologna, viale C. Berti Pichat 6/2, I-40127 Bologna, Italy}

\author{Lauro Moscardini}
\affiliation{Dipartimento di Fisica e Astronomia, Alma Mater Studiorum Universit\`a di Bologna, via Gobetti 93/2, I-40129 Bologna, Italy}
\affiliation{INAF/OAS Bologna, via Piero Gobetti 93/3, I-40129 Bologna, Italy}
\affiliation{INFN, Sezione di Bologna, viale C. Berti Pichat 6/2, I-40127 Bologna, Italy}

\author{Alfonso Veropalumbo}
\affiliation{Dipartimento di Fisica, Universit\`a degli Studi Roma Tre, via della Vasca Navale 84, I-00146 Rome, Italy}

\begin{abstract}
Features in the primordial power spectrum represent the imprinted signal in the density perturbations 
of the physics and evolution of the early Universe. A measurement of such signals will represents the 
need to go beyond the minimal assumption made for the initial conditions of the cosmological perturbations. 
For the first time, we study different templates with undamped oscillations or a bump from the two-point 
correlation function measured from BOSS DR12 galaxies constraining the amplitude of the features to be 
at most a few percent. Constraints are competitive to the ones obtained with {\em Planck} DR3.
\end{abstract}

\maketitle

\section{INTRODUCTION} \label{sec:intro}
Measurements of cosmic microwave background (CMB) anisotropies, such as those from {\em Planck} DR3 
\cite{Planck:2018vyg,Planck:2018jri,Planck:2019kim}, have contributed significantly to the 
characterization of the initial conditions of the Universe. 
The tight CMB constraints to spatial curvature, isocurvature fluctuations, and primordial non-Gaussianity, 
all agree with predictions of the standard single field slow-roll models 
\cite{Starobinsky:1980te,Guth:1980zm,Sato:1980yn,Linde:1981mu,Albrecht:1982wi,Hawking:1982ga,Linde:1983gd}.

While in these simplest realizations the nearly exponential expansion driven by a scalar field $\phi$ 
slowly rolling down a sufficiently flat potential $V(\phi)$ predicts a nearly scale-invariant primordial 
power spectrum (PPS), several classes of inflationary models predict features in the PPS 
such as scale-dependent oscillations. 

Indeed, primordial features (PF) are expected to appear as deviations from a power law in the PPS due to 
local or nonlocal modifications of the inflaton potential 
\cite{Starobinsky:1992ts,Adams:2001vc,Gong:2005jr,Chen:2006xjb,Chen:2008wn,Flauger:2009ab,Hazra:2014goa}, to the presence of new interactions and 
heavy particles \cite{Chung:1999ve,Romano:2008rr,Barnaby:2009mc,Chen:2011tu,Chen:2011zf}, non-Bunch-Davies initial 
conditions \cite{Danielsson:2002kx,Easther:2002xe,Martin:2003kp}, or also as imprints of alternatives to 
cosmic inflation \cite{Chen:2011zf,Chen:2011tu,Chen:2018cgg}. 
Moreover, phenomenologically they can be connected to persistent anomalies in the CMB temperature angular 
power spectrum measured by WMAP \cite{WMAP:2003syu} and by {\em Planck} observations 
\cite{Planck:2013jfk,Planck:2015sxf,Planck:2018jri}; PF have been also proposed to explain 
anomalies such as the CMB lensing anomaly, the $H_0$ and the $S_8$ tensions 
\cite{Planck:2018jri,Domenech:2020qay,Keeley:2020rmo,Hazra:2022rdl}.

In order to go beyond the current $2\sigma-3\sigma$ evidence of some PF patterns from CMB anisotropy 
measurements 
\cite{WMAP:2003syu,Planck:2013jfk,Meerburg:2013dla,Easther:2013kla,Chen:2014cwa,Planck:2015sxf,Planck:2018jri,Zeng:2018ufm,Braglia:2021ckn,Braglia:2021sun,Braglia:2021rej}, it has been shown how future large-scale 
structure observations will be able to improve current constraints on these oscillatory-feature models. 
This will give the opportunity to further investigate the presence of any salient feature in the matter 
power spectrum, complementing the constraints based on CMB anisotropy measurements to smaller scales, and 
being very powerful to constrain high-frequency oscillations appearing on the PPS, see 
\cite{Huang:2012mr,Chen:2016zuu,Chen:2016vvw,Ballardini:2016hpi,Slosar:2019gvt,Beutler:2019ojk,Ballardini:2019tuc}. 
In this perspective, an intensive effort has been put in place to study the nonlinear dynamics of the cold 
dark matter (CDM) power spectrum for initial conditions containing PF, see 
Refs.~\cite{Vlah:2015zda,Beutler:2019ojk,Vasudevan:2019ewf,Ballardini:2019tuc,Chen:2020ckc,Li:2021jvz}.

The goal of this work is to provide new constraints on the amplitude of PF for models with superimposed 
undamped oscillations or a bump performing the first analysis of PF models with the galaxy two-point 
correlation function (2PCF) from the Sloan Digital Sky Survey III Baryon Oscillation Spectroscopic Survey 
Data Release 12 (BOSS DR12).

Our paper is organized as follows. After this introduction, we described the primordial feature models 
and the corresponding PPS templates studied in Sec.~\ref{sec:features}. 
In Sec.~\ref{sec:data}, we described the datasets considered and the modeling for the 2PCF used for 
the analysis together with the prior on the cosmological and nuisance parameters in Sec.~\ref{sec:models}. 
We present our results obtained with SDSS BOSS DR12 in Sec.~\ref{sec:results}. 
In Sec.~\ref{sec:conclusions}, we draw our conclusions.

\section{MODELS OF OSCILLATORY PRIMORDIAL POWER SPECTRUM} \label{sec:features}
\begin{figure}
\centering
\includegraphics[width=0.48\textwidth]{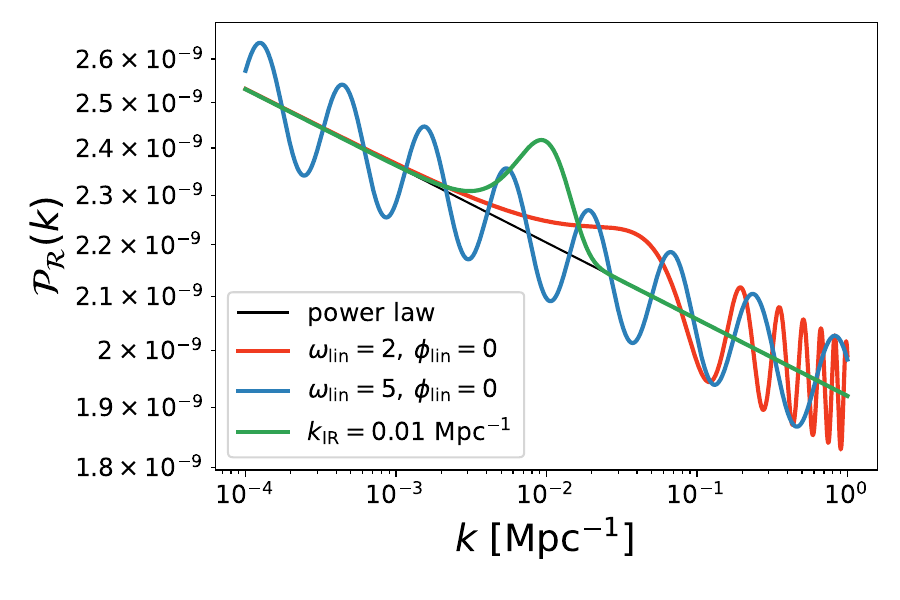}
\caption{PPS for the three templates analyzed.
For M1 (red), we fix ${\cal A}_{\rm lin} = 0.05$, $\omega_{\rm lin} = 2$, and $\phi_{\rm lin} = 0$.
For M2 (blue), we fix ${\cal A}_{\rm log} = 0.05$, $\omega_{\rm log} = 5$, and $\phi_{\rm log} = 0$.
For M3 (green), we fix $\tilde{{\cal A}}_{\rm IR} \equiv {\cal A}_{\rm IR}/A_{\rm s}= 0.1$ 
and $k_{\rm IR} = 0.01$ Mpc$^{-1}$.}
\label{fig:pk_feature}
\end{figure}
Starting from the standard power-law PPS of the comoving curvature perturbations $\cal R$ on superhorizon 
scales, which is given by
\begin{equation}
    {\cal P}_{{\cal R},\,0}(k) = A_{\rm s} \left(\frac{k}{k_*}\right)^{n_{\rm s}-1} \, ,
\end{equation}
where $A_{\rm s}$ and $n_{\rm s}$ are the amplitude and the spectral index of the comoving curvature 
perturbations at a given pivot scale $k_* = 0.05\ {\rm Mpc}^{-1}$, we consider deviations as
\begin{equation}
    {\cal P}_{\cal R}(k) = {\cal P}_{{\cal R},\,0}(k) \left[1 + {\cal P}_{\rm PF}(k)\right] \,.
\end{equation}

There are many kinds of oscillatory patterns models such as features that oscillates linearly or 
logarithmically in $k$ as well as oscillations with constant or scale-dependent amplitude 
\cite{Starobinsky:1992ts,Adams:2001vc,Danielsson:2002kx,Easther:2002xe,Martin:2003kp,Wang:2002hf,Bean:2008na,Chen:2008wn,Flauger:2009ab,Chen:2010bka,Chen:2011zf,Chen:2011tu,Chen:2018cgg}.
We therefore consider the following templates with superimposed oscillations on the PPS, 
\begin{equation} \label{eqn:pk_wigg}
    {\cal P}_{\rm PF}(k) = {\cal A}_{\rm X} \sin\left(\omega_{\rm X}\Xi_{\rm X} + 2\pi\phi_{\rm X}\right) \,,
\end{equation}
where ${\rm X} = \{{\rm lin,\,log}\}$ and $\Xi_{\rm X} = \{k/k_*,\,\log(k/k_*)\}$. These two templates 
with undamped oscillatory features (hereafter M1 and M2) are described by three extra parameters; an 
amplitude ${\cal A}_{\rm X}$, a frequency $\omega_{\rm X}$, and a phase $2\pi\phi_{\rm X}$ that we restrict 
to the range $[0,\,2\pi]$. Linear extended oscillatory features can be generated by a sharp variation of background quantities $\propto \delta(\tau-\tau_0)$ 
where $\tau_0$ determines the characteristic frequency of the oscillatory pattern. 
Such situations can be realized either by having sharp variation of the slow-roll parameters, induced for instance by a nonsmooth inflaton potential \cite{Starobinsky:1992ts,Adams:2001vc,Gong:2005jr,Chen:2006xjb,Hazra:2014goa}, or by sharp variation of the sound speed, induced for instance by a sudden turn in the multifield inflationary trajectory \cite{Achucarro:2010da,Chen:2011zf,Chen:2012ge}.
Alternatively, linear oscillatory features could arise assuming non-Bunch-Davies initial conditions; see Refs.~\cite{Danielsson:2002kx,Easther:2002xe,Martin:2003kp}.
Features with logarithmic oscillations can be generated when background quantities have an oscillatory time-dependent behavior themselves \cite{Flauger:2010ja}, 
such as models with periodic potentials \cite{Freese:1990rb,Chen:2008wn,McAllister:2008hb}. 
Note that models cold also predict a scale-dependent amplitude 
for the oscillatory signal; see Refs.~\cite{Starobinsky:1992ts,Danielsson:2002kx,Easther:2002xe,Martin:2003kp,Braglia:2021ckn}.

As third model (hereafter M3), we consider a localized bump \cite{Barnaby:2009dd}
\begin{equation} \label{eqn:pk_bump}
    {\cal P}_{\rm PF}(k) = \tilde{\cal A}_{\rm IR}\left(\frac{\pi e}{3}\right)^{3/2}
        \left(\frac{k}{k_{\rm IR}}\right)^3 e^{-\frac{\pi}{2}\left(\frac{k}{k_{\rm IR}}\right)^2}
        \left(\frac{k}{k_*}\right)^{1-n_{\rm s}} \,,
\end{equation}
described by two extra parameters; an amplitude that we rescale with the amplitude of comoving curvature 
perturbations $\tilde{{\cal A}}_{\rm IR} \equiv {\cal A}_{\rm IR}/A_{\rm s}$ and a scale $k_{\rm IR}$ 
connected with the position of the bump.
This feature is generated from particle production 
during inflation assuming a single burst of particles. 
Indeed, in models with a resonant production of particles the local interaction generates a bump in the PPS plus small linear oscillations which decay very quickly.
We show the effect of the feature parameters on ${\cal P}_{\rm PF}(k)$ in Fig.~\ref{fig:pk_feature}.

\section{DATA} \label{sec:data}
In order to constraint these models, we consider the full DR12 data set\footnote{\href{https://data.sdss.org/sas/dr12/boss/papers/clustering/}{https://data.sdss.org/sas/dr12/boss/papers/clustering/}} split into three redshift bins $0.2 < z < 0.5$, $0.5 < z < 0.75$, and (overlapping) 
$0.4 < z < 0.6$, following Ref.~\cite{Ross:2016gvb}. We study also the case combining the first two 
nonoverlapping redshift bins, neglecting their correlation.

We analyze the DR12 data set with the publicly available library 
{\tt CosmoBolognaLib}\footnote{\href{https://gitlab.com/federicomarulli/CosmoBolognaLib}{https://gitlab.com/federicomarulli/CosmoBolognaLib}} \cite{Marulli:2015jil} by performing a Markov Chain Monte Carlo (MCMC) analysis.

For all redshift bins, we use the first two even multipole moments, i.e., the monopole and the quadrupole, 
of the anisotropic correlation function $\xi(s,\mu)$ measured using the minimum-variance unbiased estimator 
\cite{Landy:1993yu} with density-field reconstruction \cite{Eisenstein:2006nk,Padmanabhan:2012hf}
\begin{equation}
    \xi(s,\mu) = \frac{DD(s,\mu)-2DS(s,\mu)+SS(s,\mu)}{SS(s,\mu)} \,,
\end{equation}
where $DD$, $SS$, and $DS$ are the numbers of galaxy-galaxy, random-random, and galaxy-random pairs 
in bins of $s$ and $\mu$.

Unless otherwise noted, we use a flat $\Lambda$CDM cosmology given by $\Omega_{\rm m} =  0.31$, 
$\Omega_{\rm b} h^2 = 0.0220$, $h = 0.676$, $n_{\rm s} = 0.97$, $\sigma_8 = 0.8$, and three massless 
neutrinos. This is consistent with {\em Planck} DR3 results \cite{Planck:2018vyg} and is the same 
cosmological model as used in the papers studying the BOSS DR12 combined sample \cite{Ross:2016gvb}.
We consider the postreconstrunced galaxy 2PCF on the range of scales $20 < s\ [h^{-1}{\rm Mpc}] < 180$ 
divided in $\Delta s = 5\ h^{-1}{\rm Mpc}$ bins.

\begin{table}
\centering
\begin{tabular}{cc}
\hline
Parameter & Flat prior range  \\
\hline
$\alpha_\perp,\,\alpha_\parallel$     & $[0.7,1.3]$  \\
$B_0,\,B_2$              & $[0,5]$  \\
$A^0_0,\,A^2_0,\,A^0_1,\,A^2_1$            & $[-100,100]$  \\
$A^0_2,\,A^2_2$            & $[-1000,1000]$  \\ 
\hline
${\cal A}_{\rm lin,\,log}$  & $[0,0.5]$  \\
$\tilde{{\cal A}}_{\rm IR}$ & $[0,1]$  \\
$\omega_{\rm lin}$          & $[1.5,13]$  \\
$\omega_{\rm log}$          & $[1,50]$  \\
$k_{\rm IR}$ [Mpc$^{-1}$]   & $[0.001,0.1]$  \\
$\phi_{\rm lin,\,log}$      & $[0,1]$  \\
\hline
\end{tabular}
\caption{Priors for model parameters.}
\label{tab:prior}
\end{table}

\section{ANISOTROPIC ANALYSIS OF THE BAO PEAK IN THE MULTIPOLES OF THE 2PCF} \label{sec:models}
To build the model for the anisotropic 2PCF, we start modeling the nonlinear CDM power spectrum in 
redshift space \cite{Kaiser:1987qv,Fisher:1993pz,Xu:2012hg,Xu:2012fw}
\begin{align}
    &P^{\rm CDM}(k,\mu) = [1+\beta\mu^2R(k,\Sigma_r)]^2F_{\rm fog}(k,\Sigma_s)P_{\rm nw}(k) \notag\\
    &\qquad\qquad\qquad \times \left[1 + P_{\rm BAO}(k)e^{-k^2\Sigma_{\rm NL}^2(\mu)/2}\right] \notag\\
    &\qquad\qquad\qquad \times \left[1 + P_{\rm PF}(k)e^{-k^2\Sigma_{\rm PF}^2/2}\right] \label{eqn:Pk_ani} \,,
\end{align}
where
\begin{align}
    &F_{\rm fog}(k,\mu,\Sigma_s) = \frac{1}{\left(1+k^2\mu^2\Sigma_s^2/2\right)^2} \,\\
    &P_{\rm BAO}(k) = \frac{P_{\rm lin}(k)}{P_{\rm nw}(k)} - 1 \,\\
    &\Sigma_{\rm NL}^2(\mu,\Sigma_\parallel,\Sigma_\perp) = \mu^2\Sigma_\parallel^2+(1-\mu^2)\Sigma_\perp^2 \,.
\end{align}
The linear power spectrum $P_{\rm lin}(k)$ is obtained with 
{\tt CAMB}\footnote{\href{https://camb.readthedocs.io/en/latest/}{https://camb.readthedocs.io/en/latest/}} 
\cite{Lewis:1999bs,Howlett:2012mh} and the {\em no-wiggle} $P_{\rm nw}(k)$ is obtained with fitting 
formulae from Ref.~\cite{Eisenstein:1997ik}, in both cases using our fiducial cosmology.
$\beta$ is the ratio between the linear growth rate $f(z)=\Omega_{\rm m}(z)^{0.545}$ and the linear clustering bias.
We fix the streaming scale $\Sigma_s = 4\ h^{-1}{\rm Mpc}$. 
The radial and transverse components of the standard Gaussian damping of baryon acoustic oscillations (BAO) 
are fixed at $\Sigma_\parallel = 4\ h^{-1}{\rm Mpc}$ and $\Sigma_\perp = 2.5\ h^{-1}{\rm Mpc}$. 
$R(k,\Sigma_r) = 1 - e^{-k^2\Sigma_r^2/2}$ is the smoothing applied in reconstruction and 
$\Sigma_r = 15\ h^{-1}{\rm Mpc}$ is the smoothing scale used when deriving the displacement field 
\cite{Seo:2015eyw}. Motivated by the weak dependence of current measurements of the 2PCF on the choice of 
damping scales (see Ref.~\cite{Ross:2016gvb}), we fix $\Sigma_{\rm NL} = \Sigma_{\rm PF}$.
We test changing the fixed value of the Gaussian damping to $2\Sigma_{\rm PF}$, 
i.e. with a larger damping of the primordial features, finding negligible effects on the final constraints.

\begin{figure*}
\centering
\includegraphics[width=0.32\textwidth]{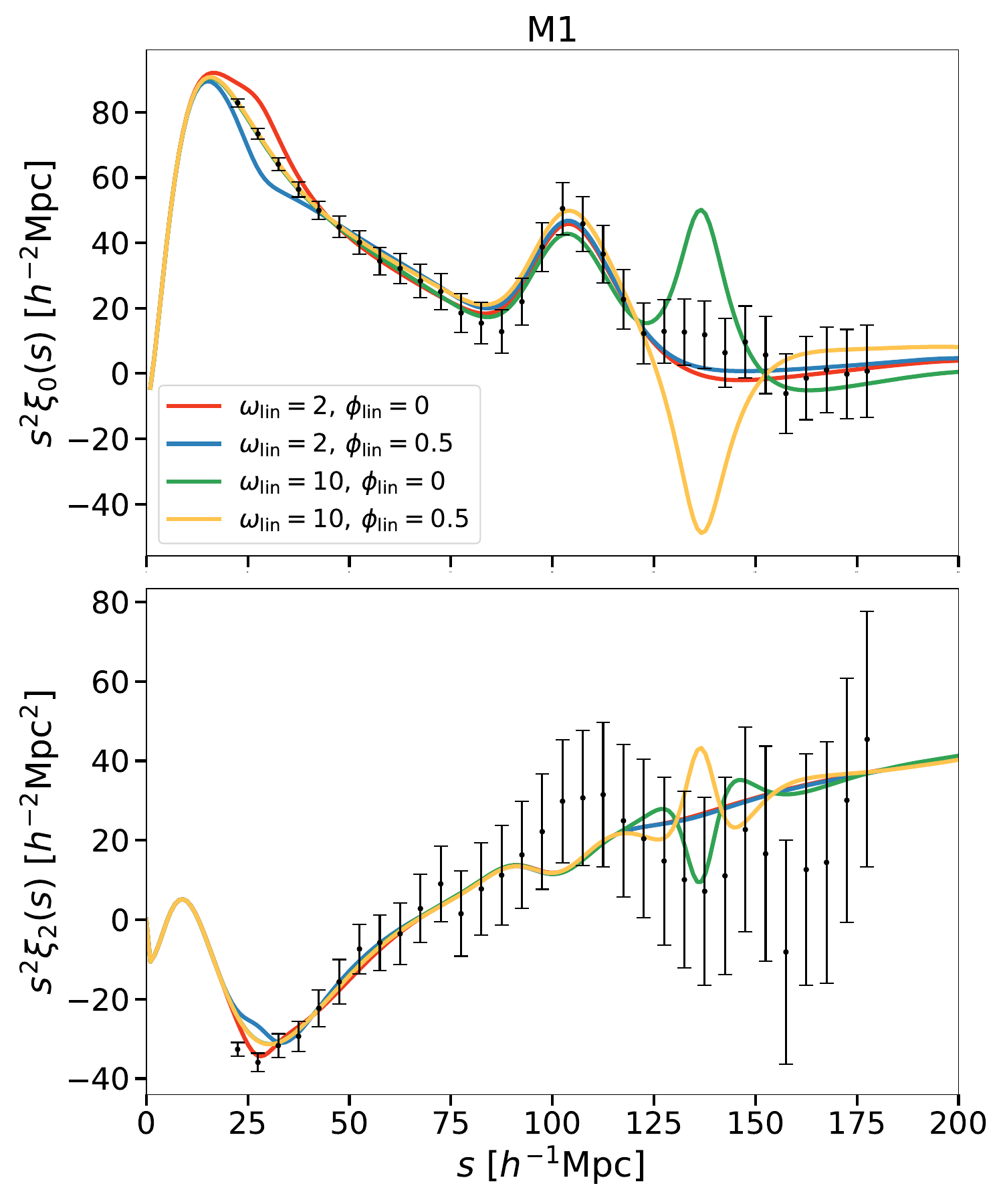}
\includegraphics[width=0.32\textwidth]{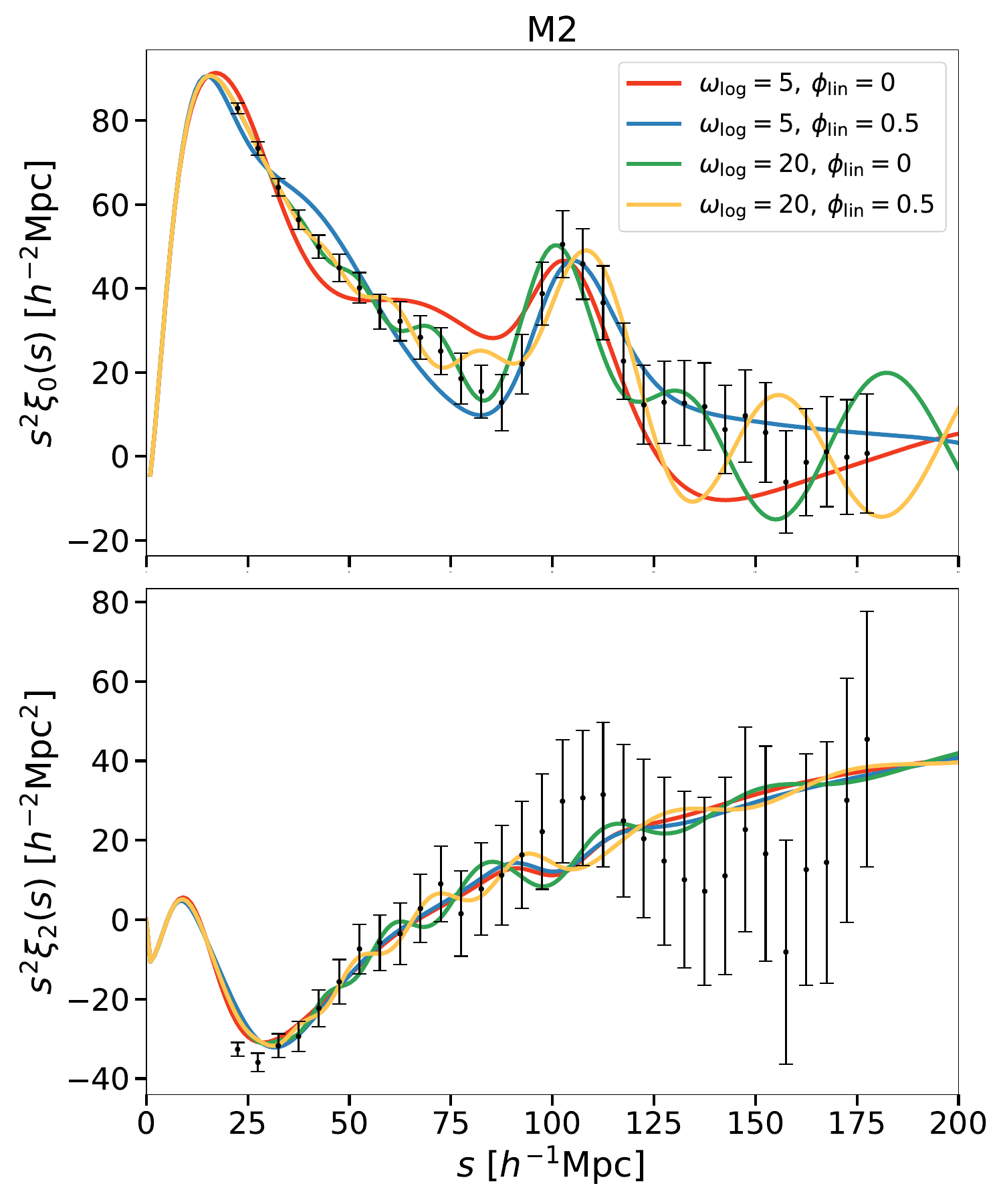}
\includegraphics[width=0.32\textwidth]{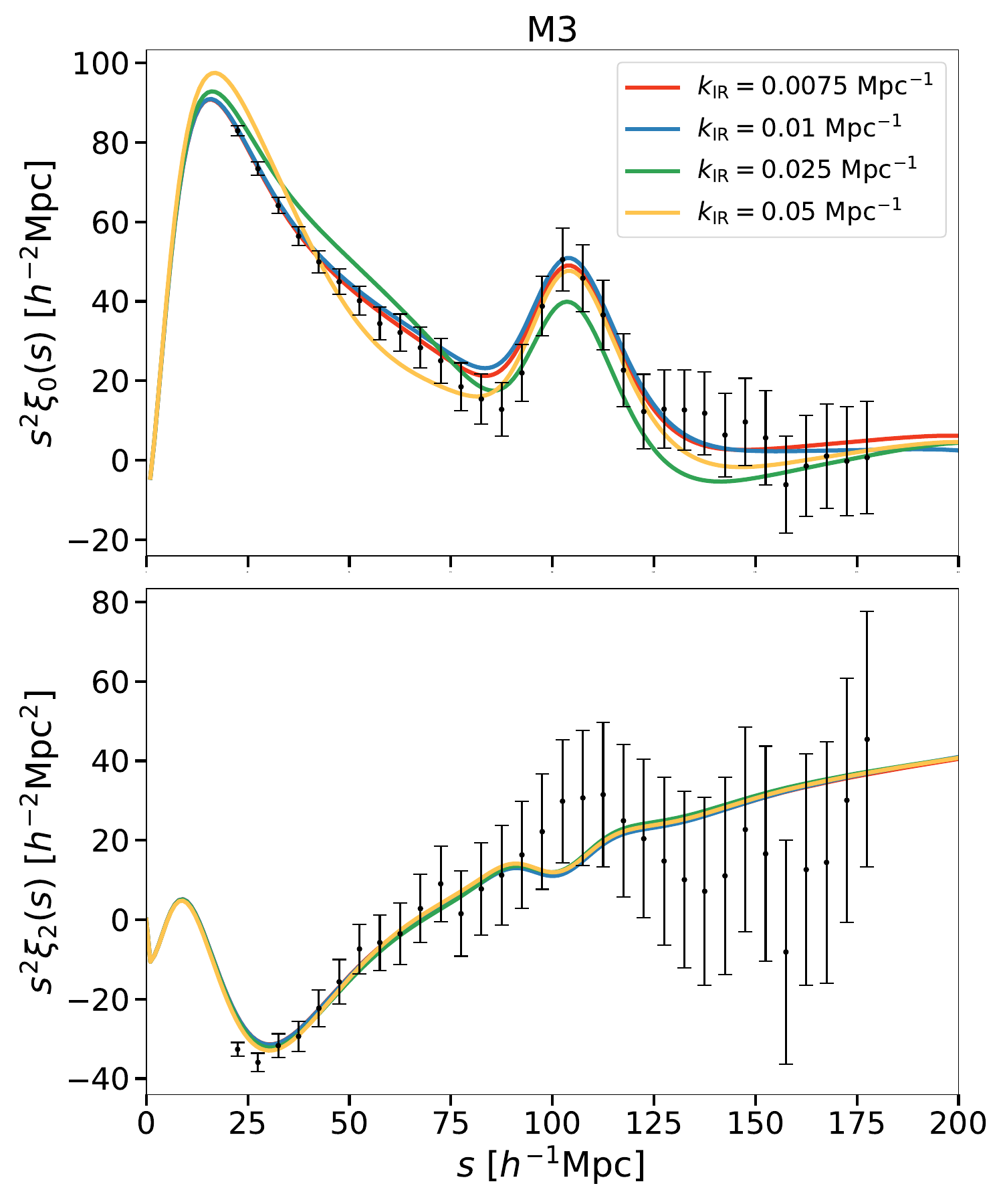}
\caption{Nonlinear galaxy 2PCF monopole (top panels) and quadrupole (bottom panels) computed at $z=0.51$ 
for M1, M2, and M3 (from left to right). We fix the standard cosmological and model parameters to the 
baseline and we vary one feature parameter at a time.
For M1 (left panel), we fix ${\cal A}_{\rm lin} = 0.05$ and we vary $\omega_{\rm lin} = \{2,\,10\}$, 
$\phi_{\rm lin} = \{0,\,0.5\}$. 
For M2 (central panel), we fix ${\cal A}_{\rm log} = 0.05$ and we vary $\omega_{\rm log} = \{5,\,20\}$, 
and $\phi_{\rm log} = \{0,\,0.5\}$.
For M3 (right panel), we fix $\tilde{{\cal A}}_{\rm IR} \equiv {\cal A}_{\rm IR}/A_{\rm s}= 0.1$ and 
we vary $k_{\rm IR} = \{0.0075,\,0.01,\,0.025,\,0.05\}$ Mpc$^{-1}$.}
\label{fig:xi_feature}
\end{figure*}

Following Ref.~\cite{Veropalumbo:2013cua,Ross:2016gvb}, we fit to the data using the following model for the monopole 
and quadrupole of the galaxy 2PCF
\begin{align}
    &\xi_0(s) = B_0\xi_0^{\rm CDM}(s,\alpha_{\perp},\alpha_{\parallel})
        + A_0^0 + \frac{A_0^1}{s} + \frac{A_0^2}{s^2}\, ,\\
    &\xi_2(s) = \frac{5}{2}\left[B_2\xi_2^{\rm CDM}(s,\alpha_{\perp},\alpha_{\parallel})
        - B_0\xi_0^{\rm CDM}(s,\alpha_{\perp},\alpha_{\parallel})\right] \notag\\
        &\qquad\qquad + A_2^0 + \frac{A_2^1}{s} + \frac{A_2^2}{s^2} \,,
\end{align}
where $\xi_0^{\rm DM}(s,\alpha_{\perp},\alpha_{\parallel})$ and 
$\xi_2^{\rm DM}(s,\alpha_{\perp},\alpha_{\parallel})$ are the CDM 2PCF monopole and quadrupole computed at 
the fiducial cosmology, and \{$B_0$, $B_2$, $A^0_0$ $A^2_0$, $A^0_1$, $A^2_1$, $A^0_2$, $A^2_2$\} are 
considered as nuisance parameters.

Given Eq.~\eqref{eqn:Pk_ani}, we calculate the templates $\xi_0^{\rm DM}(s,\alpha_{\perp},\alpha_{\parallel})$ and 
$\xi_2^{\rm DM}(s,\alpha_{\perp},\alpha_{\parallel})$ 
starting from the definition of multipole moments
\begin{equation}
    P_\ell(k) = \frac{2\ell+1}{2}\int_{-1}^{+1}{\rm d}\mu P(k,\mu)L_\ell(\mu) \, ,
\end{equation}
where $L_\ell(\mu)$ are Legendre polynomials. These are transformed to 2PCF multipoles as 
\begin{equation}
    \xi_\ell(s) = \frac{i^\ell}{2\pi^2} \int {\rm d}k\,k^2 P_\ell(k)j_\ell(ks) \, ,
\end{equation}
where $j_\ell(ks)$ is the $\ell$th order spherical Bessel function. We then use
\begin{equation}
    \xi(s,\,\mu) = \sum_{\ell=0,2} \xi_\ell(s)L_\ell(\mu) \,.
\end{equation}
Finally, we take averages over $\mu$
\begin{equation}
    \xi_\ell(s,\alpha_\perp,\alpha_\parallel) = \int_{-1}^{1}{\rm d}\mu P_\ell(\mu_{\rm true})\xi(s_{\rm true},\mu_{\rm true}) \, ,
\end{equation}
where $\mu_{\rm true} = \mu\alpha_\parallel/\sqrt{\mu^2\alpha_\parallel^2+(1-\mu^2)\alpha^2_\perp}$, 
$s_{\rm true} = s\sqrt{\mu^2\alpha_\parallel^2+(1-\mu^2)\alpha^2_\perp}$.
All the nuisance parameters, i.e. $B_0$, $B_2$ and $A^i_j$, are used 
to marginalize over the clustering bias amplitude, redshift-space distortions, and the broadband effects including angle-dependent over all shape of the power spectra.
We impose uniform flat prior on all parameters (see Table~\ref{tab:prior} for the prior ranges).

The effect of each of the feature parameters on the nonlinear galaxy 2PCF monopole and quadrupole is illustrated 
in Fig.~\ref{fig:xi_feature} for a reference central redshift $z=0.51$ of the BOSS DR12 sample.
We see that M1 present a sharp peak or deep, depending on the value of the phase $\phi_{\rm lin}$, 
corresponding to the scale of the 2PCF located at $s \sim 20\,\omega_{\rm lin}\,$Mpc. 
The signal is smoothed by nonlinear effects similarly to the case of the BAO 
\cite{Beutler:2019ojk,Ballardini:2019tuc}, damping the primordial oscillations on the small 
scales of the Fourier space matter power spectrum and leading to a broad bump on the 2PCF. Note that for 
$\omega_{\rm lin} \sim 7.5$ and $\phi_{\rm lin} = 0$ the peak coincides with the position of the BAO one, 
i.e., $s_{\rm BAO} \sim 150\,$Mpc. M2 leaves a broad pattern on all scales in the 2PCF. Finally, we see, 
differently from what happened for the oscillatory templates, M3 generates a smooth and broad change of 
the 2PCF monopole shape. It is interesting to note that while the first two templates leave a clear 
oscillatory pattern also on the quadrupole, for the third model there is a negligible effect on the 
quadrupole.

\section{RESULTS FOR THE ANISOTROPIC ANALYSIS} \label{sec:results}
\begin{table*}
\centering
\begin{tabular}{ccccccccc}
\hline
 & \multicolumn{2}{c}{$\Lambda$CDM} & \multicolumn{2}{c}{M1} & \multicolumn{2}{c}{M2} & \multicolumn{2}{c}{M3} \\
 \hline
 sample & $\alpha_\perp$ & $\alpha_\parallel$ & $\alpha_\perp$ & $\alpha_\parallel$ & $\alpha_\perp$ & $\alpha_\parallel$ & $\alpha_\perp$ & $\alpha_\parallel$ \\
\hline
$0.2 < z < 0.5$      & $0.979\pm 0.014$   & $1.017\pm 0.028$   & $0.982\pm 0.017$   & $1.004\pm 0.029$ & $0.976\pm 0.015$  & $1.022\pm 0.027$  & $0.980\pm 0.014$ & $1.015\pm 0.028$ \\
$0.4 < z < 0.6$      & $0.991\pm 0.014$   & $0.988\pm 0.025$   & $0.994\pm 0.016$   & $0.989\pm 0.027$ & $0.992\pm 0.014$  & $0.986\pm 0.025$  & $0.992\pm 0.014$ & $0.988\pm 0.025$ \\
$0.5 < z < 0.75$     & $0.994\pm 0.023$   & $0.953\pm 0.041$   & $0.998\pm 0.024$   & $0.948\pm 0.045$ & $0.995\pm 0.025$  & $0.950\pm 0.044$  & $0.992\pm 0.021$ & $0.957\pm 0.040$ \\
\hline
{\em combined} & $0.976\pm 0.014$   & $1.031\pm 0.027$   & $0.985 \pm 0.018$  & $1.017 \pm 0.027$ & $0.982\pm 0.016$  & $1.016\pm 0.028$ & $0.979\pm 0.014$ & $1.018\pm 0.029$ \\
\hline
\end{tabular}
\caption{Marginalized mean values and 68\% C.L. for the 2D BAO parameters $\alpha_\perp$ and $\alpha_\parallel$ 
for the $\Lambda$CDM, M1, M2, and M3.}
\label{tab:alpha_ani_bao}
\end{table*}
We start performing the standard BAO analysis on BOSS DR12 data for $\Lambda$CDM cosmology to validate our 
pipeline, using the measured postreconstructed monopole and quadrupole, for each redshift bin. Means and 
uncertainties for $\alpha_\perp$ and $\alpha_\parallel$, marginalized over 
\{$B_0$, $B_2$, $A^0_0$ $A^2_0$, $A^0_1$, $A^2_1$, $A^0_2$, $A^2_2$\}, agree well with the ones found in 
Ref.~\cite{Ross:2016gvb}; we collect the results in Table~\ref{tab:alpha_ani_bao}.
We repeat the BAO analysis in the presence of the template with linear oscillations (M1), logarithmic ones 
(M2), and a bump (M3). The constraints on $\alpha_\perp,\, \alpha_\parallel$ are affected in some cases by 
$< 20\%$ larger uncertainties, see Table~\ref{tab:alpha_ani_bao}.

\begin{figure}
\centering
\includegraphics[width=0.43\textwidth]{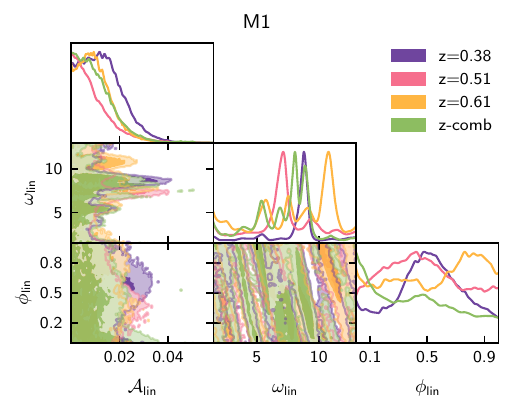}
\includegraphics[width=0.43\textwidth]{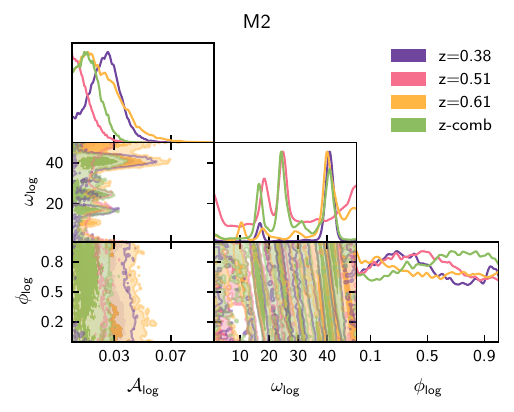}
\includegraphics[width=0.43\textwidth]{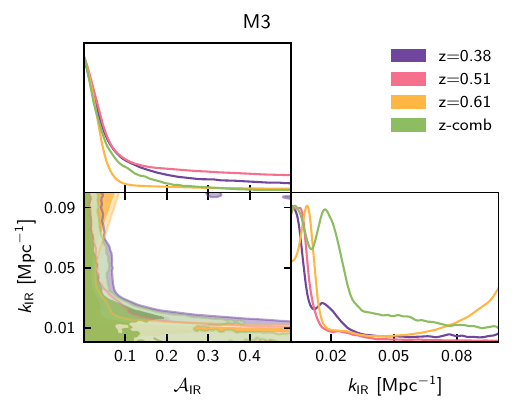}
\caption{Marginalized joint posterior distributions of the features parameters for M1, M2 and M3 
models (from top to bottom).}
\label{fig:features_ani_wigg}
\end{figure}

\begin{figure}
\centering
\includegraphics[width=0.48\textwidth]{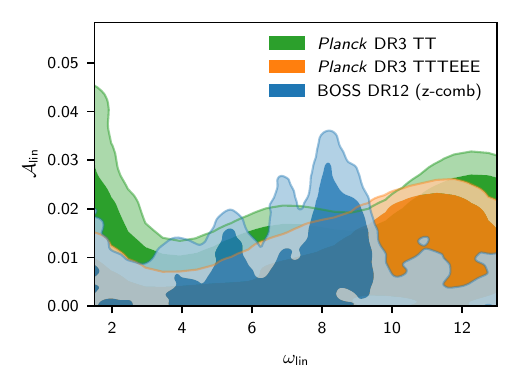}
\includegraphics[width=0.48\textwidth]{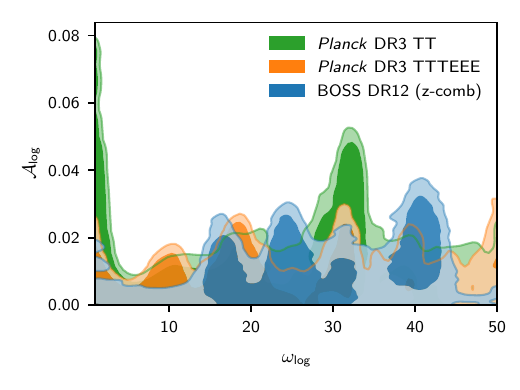}
\caption{Marginalized joint posterior distribution of the value of the feature parameters 
for M1 model (top panel) and M2 (bottom panel) obtained with the BOSS DR12 combined redshift bin 
and for the {\em Planck} DR3.}
\label{fig:features_ani_cmb}
\end{figure}

For M1, the marginalized 95\% confidence level (C.L.) upper bound on the amplitude of the primordial 
oscillations analyzing the 2PCF 
is ${\cal A}_{\rm lin} < 0.029$ at $z = 0.38$, ${\cal A}_{\rm lin} < 0.024$ at $z = 0.51$, 
${\cal A}_{\rm lin} < 0.023$ at $z = 0.61$ on the frequency range $1.5 < \omega_{\rm lin} < 13$; 
combining the two nonoverlapping samples we obtain ${\cal A}_{\rm lin} < 0.025$ at 95\% C.L.
For M2, the marginalized 95\% C.L. upper bound on the amplitude of the primordial oscillations 
is ${\cal A}_{\rm log} < 0.048$ at $z = 0.38$, 
${\cal A}_{\rm log} < 0.023$ at $z = 0.51$, ${\cal A}_{\rm log} < 0.047$ at $z = 0.61$ on the frequency 
range $1 < \omega_{\rm log} < 50$; combining the two nonoverlapping samples we obtain 
${\cal A}_{\rm log} < 0.027$ at 95\% C.L.
For M3, the marginalized 95\% C.L. upper bound on the amplitude of the bump is 
$\tilde{\cal A}_{\rm IR} < 0.88$ at $z = 0.38$, $\tilde{\cal A}_{\rm IR} < 0.92$ at $z = 0.51$, 
$\tilde{\cal A}_{\rm IR} < 0.86$ at $z = 0.61$; combining the two nonoverlapping samples we obtain 
$\tilde{\cal A}_{\rm IR} < 0.34$ at 95\% C.L. 
Constraints on the amplitude $\tilde{\cal A}_{\rm IR}$, derived using \eqref{eqn:pk_bump}, depends on the details of the underlying model, in particular on the interaction term regulating the production of particles and the presence of a single or multiple bursts of particles.
For a simple model where the inflation field $\phi$ interacts 
with the field $\chi$ through the Lagrangian term
\begin{equation}
    {\cal L} = -\frac{g^2}{2}\left(\phi-\phi_0\right)^2\chi^2 \,,
\end{equation}
we can derive constraints on the coupling $g$ thanks to the relation 
${\cal A}_{\rm IR} \simeq 1.01 \times 10^{-6} g^{15/4}$ for a single burst of particle production 
during inflation \cite{Barnaby:2009dd}.
We find for the combined sample $g^2 < 0.021$ at 95\% C.L. (for $A_{\rm s}= 2.1\times 10^{-9}$), 
corresponding to a reasonable coupling values for models with phase transition during inflation 
and comparable in magnitude to the constraint obtained from CMB data in Refs.~\cite{Barnaby:2009dd,Naik:2022mxn}. 

In Fig.~\ref{fig:features_ani_wigg}, we show the comparison of the constraints on the feature 
parameters for all the three models studied for the BOSS DR12 single redshift analysis and the combined case.

The bounds on the first two templates with oscillatory features can be compared with the results inferred 
from CMB anisotropy observations obtained with {\em Planck} DR3 for the same templates \cite{Planck:2018jri}. 
The frequency coverage in the CMB analysis is wider for both templates, i.e., $1 < \omega_X < 10^{2.1}$. 
On the restricted frequency range covered by our analysis, the constraints from the 2PCF are comparable 
if not slightly tighter compared to the CMB ones. 
{\em Planck} DR3 constraints at 95 \% C.L. correspond to ${\cal A}_{\rm lin} < 0.029$ ($< 0.028$) for TT 
(TTTEEE) for M1 and ${\cal A}_{\rm log} < 0.043$ ($< 0.023$) for TT (TTTEEE) for M2.
In Fig.~\ref{fig:features_ani_cmb}, we overplot the marginalized posterior distribution on the feature 
amplitude depending on the frequency from our BOSS DR12 analysis with the one obtained from CMB {\em Planck} 
DR3.
We are able to put tight constraints on some of the best-fit frequencies detected with CMB {\em Planck} 
data \cite{Planck:2018jri} using BOSS DR12 measurements. In particular, the best fit for M1 from TTTEEE 
data corresponding to ${\cal A}_{\rm lin} = 0.015$ and $\omega_{\rm lin} \simeq 10$, and the best-fit for 
M2 from TT data corresponding to ${\cal A}_{\rm log} = 0.024$ and $\omega_{\rm lin} \simeq 32$ are at odds 
with current LSS data. 

Our constraints for M1 and M2 are consistent with the ones obtained in \cite{Beutler:2019ojk} where the 
same two templates have been constrained with the isotropic power spectrum monopole in Fourier space on 
the frequency ranges $5 < \omega_{\rm lin} < 45$ and $10 < \omega_{\rm log} < 80$, finding bounds on the 
features amplitude from BOSS DR12 up to a factor of 2.3 and 3.1 tighter than the ones obtained from CMB 
for $\omega_{\rm lin} > 10$ and $\omega_{\rm log} > 20$, respectively.

\section{CONCLUSIONS} \label{sec:conclusions}
In this paper, we have derived for the first time constraints on primordial feature 
in the PPS through the galaxy two-point correlation function.
We have considered undamped superimposed oscillations, and a localized bump motivated by 
the production of particles during inflation.
Our results show that the BOSS DR12 2PCF has the potential to achieve sensitivities competitive 
to current limits from {\em Planck} DR3 CMB anisotropies, improving the bounds on the average 
amplitude of such features and putting at odds some of the CMB best-fit frequencies found in 
{\em Planck} data. 
The complementarity between 2PCF and CMB measurements of primordial features opens exciting 
synergies, as verification of the primordial interpretation of anomalies in the CMB temperature 
and polarization angular power spectrum, that will lead the way to new studies of the physics 
of the early Universe.

Future galaxy clustering measurements, such as those expected from DESI \cite{DESI:2016fyo}, 
Euclid \cite{EUCLID:2011zbd}, SPHEREx \cite{Dore:2014cca}, will significantly improve the current 
bounds obtained here in different aspects. 
Beyond a lower shot noise and higher redshift bins, a wider redshift volume will allow to probe 
larger separation scales and therefore target even higher frequencies for superimposed 
oscillations than those studied here.

\section*{Acknowledgments}
We all acknowledge financial support from the contract ASI/ INAF for the Euclid mission No. 2018-23-HH.0.
F.F. acknowledges support by the Agreement No. 2020-9-HH.0 ASI-UniRM2 ``Partecipazione italiana alla fase 
A della missione LiteBIRD''.
L.M. acknowledges support from the Grant No. PRIN-MIUR 2017 WSCC32.

\footnotesize
\bibliographystyle{abbrv}

\begin{thebibliography}{}

\end{thebibliography}


\begin{thebibliography}{nnn}
\bibitem{Planck:2018vyg}
N.~Aghanim \textit{et al.} [Planck Collaboration],
Astron. Astrophys. \textbf{641}, A6 (2020)
[erratum: Astron. Astrophys. \textbf{652}, C4 (2021)]
doi:10.1051/0004-6361/201833910
[arXiv:1807.06209 [astro-ph.CO]].

\bibitem{Planck:2018jri}
Y.~Akrami \textit{et al.} [Planck Collaboration],
Astron. Astrophys. \textbf{641}, A10 (2020)
doi:10.1051/0004-6361/201833887
[arXiv:1807.06211 [astro-ph.CO]].

\bibitem{Planck:2019kim}
Y.~Akrami \textit{et al.} [Planck Collaboration],
Astron. Astrophys. \textbf{641}, A9 (2020)
doi:10.1051/0004-6361/201935891
[arXiv:1905.05697 [astro-ph.CO]].

\bibitem{Starobinsky:1980te}
A.~A.~Starobinsky,
Phys. Lett. B \textbf{91}, 99-102 (1980)
doi:10.1016/0370-2693(80)90670-X

\bibitem{Guth:1980zm}
A.~H.~Guth,
Phys. Rev. D \textbf{23}, 347-356 (1981)
doi:10.1103/PhysRevD.23.347

\bibitem{Sato:1980yn}
K.~Sato,
Mon. Not. Roy. Astron. Soc. \textbf{195}, 467-479 (1981)
NORDITA-80-29.

\bibitem{Linde:1981mu}
A.~D.~Linde,
Phys. Lett. B \textbf{108}, 389-393 (1982)
doi:10.1016/0370-2693(82)91219-9

\bibitem{Albrecht:1982wi}
A.~Albrecht and P.~J.~Steinhardt,
Phys. Rev. Lett. \textbf{48}, 1220-1223 (1982)
doi:10.1103/PhysRevLett.48.1220

\bibitem{Hawking:1982ga}
S.~W.~Hawking, I.~G.~Moss and J.~M.~Stewart,
Phys. Rev. D \textbf{26}, 2681 (1982)
doi:10.1103/PhysRevD.26.2681

\bibitem{Linde:1983gd}
A.~D.~Linde,
Phys. Lett. B \textbf{129}, 177-181 (1983)
doi:10.1016/0370-2693(83)90837-7

\bibitem{Starobinsky:1992ts}
A.~A.~Starobinsky,
JETP Lett. \textbf{55}, 489-494 (1992)

\bibitem{Adams:2001vc}
J.~A.~Adams, B.~Cresswell and R.~Easther,
Phys. Rev. D \textbf{64}, 123514 (2001)
doi:10.1103/PhysRevD.64.123514
[arXiv:astro-ph/0102236 [astro-ph]].

\bibitem{Gong:2005jr}
J.~O.~Gong,
JCAP \textbf{07}, 015 (2005)
doi:10.1088/1475-7516/2005/07/015
[arXiv:astro-ph/0504383 [astro-ph]].

\bibitem{Chen:2006xjb}
X.~Chen, R.~Easther and E.~A.~Lim,
JCAP \textbf{06}, 023 (2007)
doi:10.1088/1475-7516/2007/06/023
[arXiv:astro-ph/0611645 [astro-ph]].

\bibitem{Hazra:2014goa}
D.~K.~Hazra, A.~Shafieloo, G.~F.~Smoot and A.~A.~Starobinsky,
JCAP \textbf{08}, 048 (2014)
doi:10.1088/1475-7516/2014/08/048
[arXiv:1405.2012 [astro-ph.CO]].

\bibitem{Chen:2008wn}
X.~Chen, R.~Easther and E.~A.~Lim,
JCAP \textbf{04}, 010 (2008)
doi:10.1088/1475-7516/2008/04/010
[arXiv:0801.3295 [astro-ph]].

\bibitem{Flauger:2009ab}
R.~Flauger, L.~McAllister, E.~Pajer, A.~Westphal and G.~Xu,
JCAP \textbf{06}, 009 (2010)
doi:10.1088/1475-7516/2010/06/009
[arXiv:0907.2916 [hep-th]].

\bibitem{Chung:1999ve}
D.~J.~H.~Chung, E.~W.~Kolb, A.~Riotto and I.~I.~Tkachev,
Phys. Rev. D \textbf{62}, 043508 (2000)
doi:10.1103/PhysRevD.62.043508
[arXiv:hep-ph/9910437 [hep-ph]].

\bibitem{Romano:2008rr}
A.~E.~Romano and M.~Sasaki,
Phys. Rev. D \textbf{78}, 103522 (2008)
doi:10.1103/PhysRevD.78.103522
[arXiv:0809.5142 [gr-qc]].

\bibitem{Barnaby:2009mc}
N.~Barnaby, Z.~Huang, L.~Kofman and D.~Pogosyan,
Phys. Rev. D \textbf{80}, 043501 (2009)
doi:10.1103/PhysRevD.80.043501
[arXiv:0902.0615 [hep-th]].

\bibitem{Chen:2011tu}
X.~Chen,
Phys. Lett. B \textbf{706}, 111-115 (2011)
doi:10.1016/j.physletb.2011.11.009
[arXiv:1106.1635 [astro-ph.CO]].

\bibitem{Chen:2011zf}
X.~Chen,
JCAP \textbf{01}, 038 (2012)
doi:10.1088/1475-7516/2012/01/038
[arXiv:1104.1323 [hep-th]].

\bibitem{Danielsson:2002kx}
U.~H.~Danielsson,
Phys. Rev. D \textbf{66}, 023511 (2002)
doi:10.1103/PhysRevD.66.023511
[arXiv:hep-th/0203198 [hep-th]].

\bibitem{Easther:2002xe}
R.~Easther, B.~R.~Greene, W.~H.~Kinney and G.~Shiu,
Phys. Rev. D \textbf{66}, 023518 (2002)
doi:10.1103/PhysRevD.66.023518
[arXiv:hep-th/0204129 [hep-th]].

\bibitem{Martin:2003kp}
J.~Martin and R.~Brandenberger,
Phys. Rev. D \textbf{68}, 063513 (2003)
doi:10.1103/PhysRevD.68.063513
[arXiv:hep-th/0305161 [hep-th]].

\bibitem{Chen:2018cgg}
X.~Chen, A.~Loeb and Z.~Z.~Xianyu,
Phys. Rev. Lett. \textbf{122}, no.12, 121301 (2019)
doi:10.1103/PhysRevLett.122.121301
[arXiv:1809.02603 [astro-ph.CO]].

\bibitem{WMAP:2003syu}
H.~V.~Peiris \textit{et al.} [WMAP Collaboration],
Astrophys. J. Suppl. \textbf{148}, 213-231 (2003)
doi:10.1086/377228
[arXiv:astro-ph/0302225 [astro-ph]].

\bibitem{Planck:2013jfk}
P.~A.~R.~Ade \textit{et al.} [Planck Collaboration],
Astron. Astrophys. \textbf{571}, A22 (2014)
doi:10.1051/0004-6361/201321569
[arXiv:1303.5082 [astro-ph.CO]].

\bibitem{Planck:2015sxf}
P.~A.~R.~Ade \textit{et al.} [Planck Collaboration],
Astron. Astrophys. \textbf{594}, A20 (2016)
doi:10.1051/0004-6361/201525898
[arXiv:1502.02114 [astro-ph.CO]].

\bibitem{Domenech:2020qay}
G.~Dom\`enech, X.~Chen, M.~Kamionkowski and A.~Loeb,
JCAP \textbf{10}, 005 (2020)
doi:10.1088/1475-7516/2020/10/005
[arXiv:2005.08998 [astro-ph.CO]].

\bibitem{Keeley:2020rmo}
R.~E.~Keeley, A.~Shafieloo, D.~K.~Hazra and T.~Souradeep,
JCAP \textbf{09}, 055 (2020)
doi:10.1088/1475-7516/2020/09/055
[arXiv:2006.12710 [astro-ph.CO]].

\bibitem{Hazra:2022rdl}
D.~K.~Hazra, A.~Antony and A.~Shafieloo,
JCAP \textbf{08}, no.08, 063 (2022)
doi:10.1088/1475-7516/2022/08/063
[arXiv:2201.12000 [astro-ph.CO]].

\bibitem{Meerburg:2013dla}
P.~D.~Meerburg, D.~N.~Spergel and B.~D.~Wandelt,
Phys. Rev. D \textbf{89}, no.6, 063537 (2014)
doi:10.1103/PhysRevD.89.063537
[arXiv:1308.3705 [astro-ph.CO]].

\bibitem{Easther:2013kla}
R.~Easther and R.~Flauger,
JCAP \textbf{02}, 037 (2014)
doi:10.1088/1475-7516/2014/02/037
[arXiv:1308.3736 [astro-ph.CO]].

\bibitem{Chen:2014cwa}
X.~Chen, M.~H.~Namjoo and Y.~Wang,
JCAP \textbf{02}, 027 (2015)
doi:10.1088/1475-7516/2015/02/027
[arXiv:1411.2349 [astro-ph.CO]].

\bibitem{Zeng:2018ufm}
C.~Zeng, E.~D.~Kovetz, X.~Chen, Y.~Gong, J.~B.~Mu\~noz and M.~Kamionkowski,
Phys. Rev. D \textbf{99}, no.4, 043517 (2019)
doi:10.1103/PhysRevD.99.043517
[arXiv:1812.05105 [astro-ph.CO]].

\bibitem{Braglia:2021ckn}
M.~Braglia, X.~Chen and D.~K.~Hazra,
JCAP \textbf{06}, 005 (2021)
doi:10.1088/1475-7516/2021/06/005
[arXiv:2103.03025 [astro-ph.CO]].

\bibitem{Braglia:2021sun}
M.~Braglia, X.~Chen and D.~K.~Hazra,
Eur. Phys. J. C \textbf{82}, no.5, 498 (2022)
doi:10.1140/epjc/s10052-022-10461-3
[arXiv:2106.07546 [astro-ph.CO]].

\bibitem{Braglia:2021rej}
M.~Braglia, X.~Chen and D.~K.~Hazra,
Phys. Rev. D \textbf{105}, no.10, 103523 (2022)
doi:10.1103/PhysRevD.105.103523
[arXiv:2108.10110 [astro-ph.CO]].

\bibitem{Huang:2012mr}
Z.~Huang, L.~Verde and F.~Vernizzi,
JCAP \textbf{04}, 005 (2012)
doi:10.1088/1475-7516/2012/04/005
[arXiv:1201.5955 [astro-ph.CO]].

\bibitem{Chen:2016zuu}
X.~Chen, P.~D.~Meerburg and M.~M\"unchmeyer,
JCAP \textbf{09}, 023 (2016)
doi:10.1088/1475-7516/2016/09/023
[arXiv:1605.09364 [astro-ph.CO]].

\bibitem{Chen:2016vvw}
X.~Chen, C.~Dvorkin, Z.~Huang, M.~H.~Namjoo and L.~Verde,
JCAP \textbf{11}, 014 (2016)
doi:10.1088/1475-7516/2016/11/014
[arXiv:1605.09365 [astro-ph.CO]].

\bibitem{Ballardini:2016hpi}
M.~Ballardini, F.~Finelli, C.~Fedeli and L.~Moscardini,
JCAP \textbf{10}, 041 (2016)
[erratum: JCAP \textbf{04}, E01 (2018)]
doi:10.1088/1475-7516/2016/10/041
[arXiv:1606.03747 [astro-ph.CO]].

\bibitem{Slosar:2019gvt}
A.~Slosar, K.~N.~Abazajian, M.~Abidi, P.~Adshead, Z.~Ahmed, D.~Alonso, M.~A.~Amin, B.~Ansarinejad, R.~Armstrong and C.~Baccigalupi, \textit{et al.}
Bull. Am. Astron. Soc. \textbf{51}, no.3, 98 (2019)
[arXiv:1903.09883 [astro-ph.CO]].

\bibitem{Beutler:2019ojk}
F.~Beutler, M.~Biagetti, D.~Green, A.~Slosar and B.~Wallisch,
Phys. Rev. Res. \textbf{1}, no.3, 033209 (2019)
doi:10.1103/PhysRevResearch.1.033209
[arXiv:1906.08758 [astro-ph.CO]].

\bibitem{Ballardini:2019tuc}
M.~Ballardini, R.~Murgia, M.~Baldi, F.~Finelli and M.~Viel,
JCAP \textbf{04}, no.04, 030 (2020)
doi:10.1088/1475-7516/2020/04/030
[arXiv:1912.12499 [astro-ph.CO]].

\bibitem{Vlah:2015zda}
Z.~Vlah, U.~Seljak, M.~Y.~Chu and Y.~Feng,
JCAP \textbf{03}, 057 (2016)
doi:10.1088/1475-7516/2016/03/057
[arXiv:1509.02120 [astro-ph.CO]].

\bibitem{Vasudevan:2019ewf}
A.~Vasudevan, M.~M.~Ivanov, S.~Sibiryakov and J.~Lesgourgues,
JCAP \textbf{09}, 037 (2019)
doi:10.1088/1475-7516/2019/09/037
[arXiv:1906.08697 [astro-ph.CO]].

\bibitem{Chen:2020ckc}
S.~F.~Chen, Z.~Vlah and M.~White,
JCAP \textbf{11}, 035 (2020)
doi:10.1088/1475-7516/2020/11/035
[arXiv:2007.00704 [astro-ph.CO]].

\bibitem{Li:2021jvz}
Y.~Li, H.~M.~Zhu and B.~Li,
Mon. Not. Roy. Astron. Soc. \textbf{514}, no.3, 4363-4378 (2022)
doi:10.1093/mnras/stac1544
[arXiv:2102.09007 [astro-ph.CO]].

\bibitem{Wang:2002hf}
X.~Wang, B.~Feng, M.~Li, X.~L.~Chen and X.~Zhang,
Int. J. Mod. Phys. D \textbf{14}, 1347 (2005)
doi:10.1142/S0218271805006985
[arXiv:astro-ph/0209242 [astro-ph]].

\bibitem{Bean:2008na}
R.~Bean, X.~Chen, G.~Hailu, S.~H.~H.~Tye and J.~Xu,
JCAP \textbf{03}, 026 (2008)
doi:10.1088/1475-7516/2008/03/026
[arXiv:0802.0491 [hep-th]].

\bibitem{Chen:2010bka}
X.~Chen,
JCAP \textbf{12}, 003 (2010)
doi:10.1088/1475-7516/2010/12/003
[arXiv:1008.2485 [hep-th]].

\bibitem{Achucarro:2010da}
A.~Achucarro, J.~O.~Gong, S.~Hardeman, G.~A.~Palma and S.~P.~Patil,
JCAP \textbf{01}, 030 (2011)
doi:10.1088/1475-7516/2011/01/030
[arXiv:1010.3693 [hep-ph]].

\bibitem{Chen:2012ge}
X.~Chen and Y.~Wang,
JCAP \textbf{09}, 021 (2012)
doi:10.1088/1475-7516/2012/09/021
[arXiv:1205.0160 [hep-th]].

\bibitem{Flauger:2010ja}
R.~Flauger and E.~Pajer,
JCAP \textbf{01}, 017 (2011)
doi:10.1088/1475-7516/2011/01/017
[arXiv:1002.0833 [hep-th]].

\bibitem{Freese:1990rb}
K.~Freese, J.~A.~Frieman and A.~V.~Olinto,
Phys. Rev. Lett. \textbf{65}, 3233-3236 (1990)
doi:10.1103/PhysRevLett.65.3233

\bibitem{McAllister:2008hb}
L.~McAllister, E.~Silverstein and A.~Westphal,
Phys. Rev. D \textbf{82}, 046003 (2010)
doi:10.1103/PhysRevD.82.046003
[arXiv:0808.0706 [hep-th]].

\bibitem{Barnaby:2009dd}
N.~Barnaby and Z.~Huang,
Phys. Rev. D \textbf{80}, 126018 (2009)
doi:10.1103/PhysRevD.80.126018
[arXiv:0909.0751 [astro-ph.CO]].

\bibitem{Ross:2016gvb}
A.~J.~Ross \textit{et al.} [BOSS Collaboration],
Mon. Not. Roy. Astron. Soc. \textbf{464}, no.1, 1168-1191 (2017)
doi:10.1093/mnras/stw2372
[arXiv:1607.03145 [astro-ph.CO]].

\bibitem{Marulli:2015jil}
F.~Marulli, A.~Veropalumbo and M.~Moresco,
Astron. Comput. \textbf{14}, 35-42 (2016)
doi:10.1016/j.ascom.2016.01.005
[arXiv:1511.00012 [astro-ph.CO]].

\bibitem{Landy:1993yu}
S.~D.~Landy and A.~S.~Szalay,
Astrophys. J. \textbf{412}, 64 (1993)
doi:10.1086/172900

\bibitem{Eisenstein:2006nk}
D.~J.~Eisenstein, H.~j.~Seo, E.~Sirko and D.~Spergel,
Astrophys. J. \textbf{664}, 675-679 (2007)
doi:10.1086/518712
[arXiv:astro-ph/0604362 [astro-ph]].

\bibitem{Padmanabhan:2012hf}
N.~Padmanabhan, X.~Xu, D.~J.~Eisenstein, R.~Scalzo, A.~J.~Cuesta, K.~T.~Mehta and E.~Kazin,
Mon. Not. Roy. Astron. Soc. \textbf{427}, no.3, 2132-2145 (2012)
doi:10.1111/j.1365-2966.2012.21888.x
[arXiv:1202.0090 [astro-ph.CO]].

\bibitem{Kaiser:1987qv}
N.~Kaiser,
Mon. Not. Roy. Astron. Soc. \textbf{227}, 1-27 (1987)

\bibitem{Fisher:1993pz}
K.~B.~Fisher, C.~A.~Scharf and O.~Lahav,
Mon. Not. Roy. Astron. Soc. \textbf{266}, 219-226 (1994)
doi:10.1093/mnras/266.1.219
[arXiv:astro-ph/9309027 [astro-ph]].

\bibitem{Xu:2012hg}
X.~Xu, N.~Padmanabhan, D.~J.~Eisenstein, K.~T.~Mehta and A.~J.~Cuesta,
Mon. Not. Roy. Astron. Soc. \textbf{427}, 2146 (2012)
doi:10.1111/j.1365-2966.2012.21573.x
[arXiv:1202.0091 [astro-ph.CO]].

\bibitem{Xu:2012fw}
X.~Xu, A.~J.~Cuesta, N.~Padmanabhan, D.~J.~Eisenstein and C.~K.~McBride,
Mon. Not. Roy. Astron. Soc. \textbf{431}, 2834 (2013)
doi:10.1093/mnras/stt379
[arXiv:1206.6732 [astro-ph.CO]].

\bibitem{Lewis:1999bs}
A.~Lewis, A.~Challinor and A.~Lasenby,
Astrophys. J. \textbf{538}, 473-476 (2000)
doi:10.1086/309179
[arXiv:astro-ph/9911177 [astro-ph]].

\bibitem{Howlett:2012mh}
C.~Howlett, A.~Lewis, A.~Hall and A.~Challinor,
JCAP \textbf{04}, 027 (2012)
doi:10.1088/1475-7516/2012/04/027
[arXiv:1201.3654 [astro-ph.CO]].

\bibitem{Eisenstein:1997ik}
D.~J.~Eisenstein and W.~Hu,
Astrophys. J. \textbf{496}, 605 (1998)
doi:10.1086/305424
[arXiv:astro-ph/9709112 [astro-ph]].

\bibitem{Seo:2015eyw}
H.~J.~Seo, F.~Beutler, A.~J.~Ross and S.~Saito,
Mon. Not. Roy. Astron. Soc. \textbf{460}, no.3, 2453-2471 (2016)
doi:10.1093/mnras/stw1138
[arXiv:1511.00663 [astro-ph.CO]].

\bibitem{Veropalumbo:2013cua}
A.~Veropalumbo, F.~Marulli, L.~Moscardini, M.~Moresco and A.~Cimatti,
Mon. Not. Roy. Astron. Soc. \textbf{442}, no.4, 3275-3283 (2014)
doi:10.1093/mnras/stu1050
[arXiv:1311.5895 [astro-ph.CO]].

\bibitem{Naik:2022mxn}
S.~S.~Naik, K.~Furuuchi and P.~Chingangbam,
JCAP \textbf{07}, no.07, 016 (2022)
doi:10.1088/1475-7516/2022/07/016
[arXiv:2202.05862 [astro-ph.CO]].

\bibitem{DESI:2016fyo}
A.~Aghamousa \textit{et al.} [DESI Collaboration],
[arXiv:1611.00036 [astro-ph.IM]].

\bibitem{EUCLID:2011zbd}
R.~Laureijs \textit{et al.} [Euclid Collaboration],
[arXiv:1110.3193 [astro-ph.CO]].

\bibitem{Dore:2014cca}
O.~Dor\'e, J.~Bock, P.~Capak, R.~de Putter, T.~Eifler, C.~Hirata, P.~Korngut, E.~Krause, D.~Masters and A.~Raccanelli, \textit{et al.} [SPHEREX Collaboration]
[arXiv:1412.4872 [astro-ph.CO]].

\end{thebibliography}

\end{document}